\documentclass[fontsize=12pt]{article}

\usepackage[english]{babel}
\usepackage[letterpaper,margin=1.25in]{geometry}
\usepackage{comment}

\usepackage{amsmath}
\DeclareMathOperator*{\argmax}{arg\,max}
\DeclareMathOperator*{\argmin}{arg\,min}
\usepackage{graphicx}
\usepackage[colorinlistoftodos]{todonotes}
\usepackage[colorlinks=true, allcolors=blue]{hyperref}
\usepackage{amssymb, amsfonts, amsthm}
\usepackage{graphicx}
\usepackage[round]{natbib}
\usepackage{color}
\usepackage{bm}
\usepackage{bbm}
\usepackage{subfigure}
\usepackage{graphicx}
\usepackage{mathabx}
\usepackage{multirow}
\usepackage{setspace}
\usepackage{authblk}
\usepackage{enumerate}
\usepackage{algorithm}
\usepackage{algpseudocode}
\usepackage{siunitx}
\usepackage{booktabs}
\usepackage{tabularray}
\usepackage{pdflscape}
\usepackage{afterpage}
\usepackage{tikz}
\newcommand*\circled[1]{\tikz[baseline=(char.base)]{
            \node[shape=circle,draw,inner sep=2pt] (char) {#1};}}

\newtheorem{theorem}{Theorem}
\newtheorem{lemma}[theorem]{Lemma}
\newtheorem{proposition}[theorem]{Proposition}
\newtheorem{corollary}[theorem]{Corollary}

\title{Optimal $F$-score Clustering for Bipartite Record Linkage}

\author{Eric A. Bai}
\author{Olivier Binette}
\author{Jerome P. Reiter}

\affil{Statistical Science Department, Duke University}

\begin{document}
\maketitle

\begin{abstract}
Probabilistic record linkage is often used to match records from two files, in particular when the variables common to both files comprise imperfectly measured identifiers like names and demographic variables. We consider bipartite record linkage settings in which each entity appears at most once within a file, i.e., there are no duplicates within the files, but some entities appear in both files. In this setting, the analyst desires a point estimate of the linkage structure that matches each record to at most one record from the other file. We propose an approach for obtaining this point estimate by maximizing the expected $F$-score for the linkage structure. We target the approach for record linkage methods that produce either (an approximate) posterior distribution of the unknown linkage structure or probabilities of matches for record pairs.  Using simulations and applications with genuine data, we illustrate that the $F$-score estimators can lead to sensible estimates of the linkage structure.
\end{abstract}


\section{Introduction}


Often entity resolution (ER) applications involve identifying duplicate records across two databases, where each database has no duplication within. This particular case of ER is called bipartite record linkage, since it corresponds to identifying a bipartite matching between records of the two databases. Bipartite record linkage is used for many tasks including, for example, matching a census to a post-enumeration survey for coverage estimation \citep{Jaro1989, winkler1991application}, linking conflict records for casualty estimation \citep{Sadinle2017}, public health and social sciences research \citep{jutte2011administrative}, and combining customer records from two lists in a customer relationship management system \citep{dyche2006customer}. 

Bipartite record linkage can be challenging for a variety of reasons, 
including the computational complexity associated with the comparison of record pairs, bipartite matching and transitivity constraints, and  the probabilistic dependencies that these constraints induce in the matching problem. As a result, researchers have proposed many methods for bipartite record linkage, ranging from rule-based approaches to the use of unsupervised clustering models, probabilistic classifiers, and other  algorithms supporting different steps of the linkage process \citep{christophides2020overview, papadakis2022bipartite}. For instance, \cite{Jaro1989} proposed the use of the \cite{fellegi:sunter} framework to obtain matching weights under maximum likelihood estimation, and to ensure a bipartite linkage by solving a linear sum assignment problem. This approach was improved in \cite{mcveigh2019scaling} to incorporate the bipartite linkage constraint directly into a maximum likelihood estimator. \cite{Sadinle2017} proposed a Bayesian  model that incorporates the bipartite matching constraint in a prior distribution, resulting in a Bayes linkage estimate under a chosen loss function.

Regardless of the approach, it is desirable to have a point estimate of the bipartite linkage structure.  However, existing approaches have some limitations.  
To compute a Bayes estimate, analysts need to specify a loss function, which may be difficult to align with practical requirements. For instance, default parameters of the loss function provided by \cite{Sadinle2017} can lead to a linkage that under-estimates the overlap between databases. This causes issues in applications such as casualty estimation, census coverage estimation, and other forms of population size estimation. Similar challenges arise with the use of maximum likelihood estimation, which requires analysts to tune parameters to estimate the number of matches.

To facilitate the use of bipartite record linkage algorithms, we propose a post-processing step that aligns with a commonly-used objective: to optimize the expected $F$-score of the resulting linkage under the bipartite record linkage constraint. We consider the $F$-score, the harmonic mean between precision and recall, as it is arguably the most widely-used evaluation measure in record linkage applications. As a result, optimizing the $F$-score often aligns with practical requirements.

More precisely, suppose a record linkage algorithm results in pairwise match probabilities or in a posterior distribution on the linkage structure. From those, an expected $F$-score can be computed for any possible bipartite record linkage. The optimal $F$-score clustering is the linkage that maximizes this expected $F$-score. For Bayesian record linkage, this corresponds to a Bayes estimator. When using pairwise match probabilities, this maximizes a plug-in $F$-score estimate as described in Section \ref{sec:exact}. 

Our contributions are as follows.
\begin{enumerate}
    \item We introduce an efficient algorithm to approximate the optimal $F$-score under the bipartite record linkage constraint, and use it to construct a point estimate of the linkage structure.
    \item We demonstrate that this algorithm can be adapted to any probabilistic bipartite record linkage model.
    \item We validate the approach using both simulated and genuine data, demonstrating that the $F$-score algorithm can yield more accurate point estimates of the linkage structure than existing point estimators.
    \item We discuss accurately estimating the size of the population that overlaps in the two data files.
\end{enumerate}

The remainder of the article is organized as follows. In Section \ref{sec:F-score-algo}, we introduce the optimal F-Score algorithm.  In  Section \ref{sec:estimation-overlap}, we discuss estimating the overlap population size. In Section \ref{sec:sim-test}, we investigate the performance of the algorithm using simulated and genuine data.  Finally, in Section \ref{sec:discussion}, we summarize the findings and discuss potential directions for future research.

\section{$F$-score Optimization Under a Bipartite Record Linkage Constraint}\label{sec:F-score-algo}

Let $\mathcal{A}$ and $\mathcal{B}$ be two data sets comprising  $n_{\mathcal{A}}$ and $n_{\mathcal{B}}$ records, respectively.  We presume  $n_{\mathcal{A}} \geq n_{\mathcal{B}}$, without loss of generality. For $i=1, \dots, n_{\mathcal{A}}$, each record  $\mathcal{A}_i \in \mathcal{A}$ has a unique identifier, denoted $\mathcal{A}_{i0}$.  Similarly, for $j=1, \dots, n_{\mathcal{B}}$, let $\mathcal{B}_{j0}$ denote the unique identifier for record $\mathcal{B}_{j} \in \mathcal{B}$. The unique identifiers are not observed in practice. Records $\mathcal{A}_{i}$ and $\mathcal{B}_{j}$ are called a match or a link whenever $\mathcal{A}_{i0} = \mathcal{B}_{j0}$, and called a non-match or non-link otherwise.  The goal of record linkage is to identify the record pairs $(\mathcal{A}_{i},\mathcal{B}_{j})$ with   $\mathcal{A}_{i0} = \mathcal{B}_{j0}$.

For $i = 1, \dots, n_{\mathcal{A}}$ and $j = 1, \dots, n_{\mathcal{B}}$, let $c_{i,j}=1$  when $\mathcal{A}_i$ and $\mathcal{B}_j$ are a match, and let $c_{i,j} = 0$ otherwise. Let $\mathbf{C} = [c_{i,j}]$ be the $n_{\mathcal{A}} \times n_{\mathcal{B}}$ matrix with $c_{i,j}$ as the element in the $i$th row and $j$th column. We refer to $\mathbf{C}$ as 
the {linkage structure}.  In {bipartite record linkage}, we assume that each $\mathcal{A}_i$ is linked to at most one $\mathcal{B}_j$, and vice versa. That is, for any given indices $i$ and $j$,
\begin{equation}
    \sum_{j' = 1}^{n_{\mathcal{B}}} c_{i, j'} \leq 1, \quad \sum_{i' = 1}^{n_{\mathcal{A}}} c_{i', j} \leq 1.
\end{equation}

In practice, $\mathbf{C}$ is an unknown parameter estimated through some record linkage model.  Here, we consider record linkage models that result in either a probability distribution for $\mathbf{C}$, e.g., via a Bayesian model \citep[as in, e.g., ][]{fortini2001bayesian, Tancredi2011, gutman, steortshall, Sadinle2017,  dalzellreiter, tangetal, betancourt:rodriguez, guhacausal} or marginal probabilities $p_{i,j}$ that record pairs $(\mathcal{A}_i, \mathcal{B}_j)$ are matches, e.g., from a \citet{fellegi:sunter} model as in \cite{enamorado_using_2019}. With Bayesian record linkage models, the posterior distribution for $\mathbf{C}$ is approximated usually with some Markov chain Monte Carlo sampler, resulting in $L$ plausible draws $\{\mathbf{C}^{(s)}: s = 1, \dots, L\}$ of the linkage structure. These draws can be used to compute Monte Carlo estimates of expectations or other summaries.


The posterior distribution for $\mathbf{C}$ or the estimated match probabilities generally derive from a specified probabilistic model. For now, we do not specify any particular model, since our approach to $F$-score optimization is agnostic to the structure of the model.

\subsection{$F$-score Objective Function and Estimators}

For any record pair $(\mathcal{A}_i, \mathcal{B}_j)$, let  $\hat c_{i,j}$ be an estimate of $c_{i,j}$ and $\hat{\mathbf{C}} = [\hat c_{i,j}]$ be the corresponding estimate of $\mathbf{C}$. For any $\hat{\mathbf{C}}$ and some weight $\beta > 0$, the $F$-score is defined as
\begin{equation}\label{eq:f-score}
    F_\beta(\hat{\mathbf{C}}, \mathbf{C}) = \frac{(1+\beta^2) \sum_{i,j} \hat c_{i,j} c_{i,j}}{\beta^2 \sum_{i,j} c_{i,j} +  \sum_{i,j} \hat  c_{i,j} }.
\end{equation}
The expression in \eqref{eq:f-score} represents the weighted harmonic mean between precision and recall. When $\beta = 1$, the $F$-score equally weights precision and recall. 

We propose to utilize \eqref{eq:f-score} to determine a point estimator $\hat{\mathbf{C}}_{opt}$ of $\mathbf{C}$. Specifically, we seek the $\hat{\mathbf{C}}$ that maximizes \eqref{eq:f-score}, that is,  
\begin{equation}\label{eq:opt}
    \hat{\mathbf{C}}_{\text{opt}} = \argmax_{\hat{\mathbf{C}}\in \mathcal{C}} F_\beta(\hat{\mathbf{C}}, \mathbf{C}),
\end{equation}
where $\mathcal{C}$ is the set of linkage structures satisfying the bipartite linkage condition.
Of course, typically $F_\beta(\hat{\mathbf{C}}, \mathbf{C})$ is not  observable since it requires knowledge of $\mathbf{C}$.  We therefore must estimate $F_\beta(\hat{\mathbf{C}}, \mathbf{C})$ for any $\hat{\mathbf{C}}$ under consideration. 

We consider two approaches for finding $\hat{\mathbf{C}}_{\text{opt}}$ via approximations to \eqref{eq:f-score}. The first method is appropriate for settings where a posterior distribution of $\mathbf{C}$ is available. We use the posterior distribution to approximate \eqref{eq:f-score} 
and subsequently obtain a Bayes estimate from \eqref{eq:opt}. The second method is suitable for scenarios where we have point estimates for the probability of a link for each record pair, which we denote as $p_{i,j}$. 
In this context, we approximate \eqref{eq:f-score} using a plug-in approach and obtain a point estimate by solving \eqref{eq:opt}.  We now describe these two approaches, starting with the first method.

\subsubsection{Bayes Estimator}

When a probability distribution for $\mathbf{C}$ is available, the expectation of \eqref{eq:f-score} can serve as a score function that we maximize. This yields the Bayes estimator,
\begin{equation}\label{eq:c-bayes}
    \hat{\mathbf{C}}_{\text{Bayes}} = \argmax_{\hat{\mathbf{C}}\in \mathcal{C}} \mathbb{E}\left[ F_\beta(\hat{\mathbf{C}}, \mathbf{C}) \right].
\end{equation}
Here, 
the expectation is taken with respect to the random variable $\mathbf{C}$. A closed-form expression for $\hat{\mathbf{C}}_{\text{Bayes}}$ does not exist. In Section \ref{sec:f-score-algo-detail}, we  present an optimization algorithm  to approximate the solution to \eqref{eq:c-bayes}.


\subsubsection{Optimal Score Estimator}

When estimates $\hat{p}_{i,j}$ for record pairs' $p_{i,j}$ are available, we can obtain an estimator by maximizing a plug-in estimate of the $F$-score in \eqref{eq:f-score}. That is, we define the  optimal score estimator, 
\begin{equation}\label{eq:c-ml}
    \hat{\mathbf{C}}_{\text{OS}} = \argmax_{\hat{\mathbf{C}} \in \mathcal{C}} \frac{(1+\beta^2) \sum_{i,j} \hat c_{i,j} \hat p_{i,j}}{\beta^2 \sum_{i,j} \hat p_{i,j} +  \sum_{i,j} \hat  c_{i,j} }.
\end{equation}
As with \eqref{eq:c-bayes}, no closed-form expression exists for $\hat{\mathbf{C}}_{\text{OS}}$. However, the algorithm in Section \ref{sec:f-score-algo-detail} provides an exact solution to \eqref{eq:c-ml}.


\subsection{Algorithm for Approximating the $F$-score}\label{sec:f-score-algo-detail}



We first present the algorithm for approximating $\hat{\mathbf{C}}_{\text{Bayes}}$ from \eqref{eq:c-bayes}. This algorithm is sufficiently general to be adapted for determining $\hat{\mathbf{C}}_{\text{OS}}$ from \eqref{eq:c-ml}, which we discuss subsequently.

\subsubsection{An Approximation of $\hat{\mathbf{C}}_{\text{Bayes}}$}\label{sec:approx}

To approximate $\hat{\mathbf{C}}_{\text{Bayes}}$ from \eqref{eq:c-bayes}, we adopt the general framework of outer and inner maximization proposed by \citet{jansche-2007-maximum}. Speaking broadly, for every possible number of matches $k$ within the range $0 \leq k \leq n_\mathcal{B}$, we perform an inner maximization. This step involves approximating the $\hat{\mathbf{C}}$ that optimizes \eqref{eq:f-score} for a given $k$, which we denote as $\hat{\mathbf{C}}_{\text{Bayes}}(k)$. Then, in the outer maximization step, we search across all feasible values of $k$. The $\hat{\mathbf{C}}_{\text{Bayes}}(k)$ that yields the highest value of \eqref{eq:f-score} is the point estimator for $\mathbf{C}$.

More formally, for a given $k$, let
\begin{align}
     \hat{\mathbf{C}}_{\text{Bayes}}(k) &= \argmax_{\hat{\mathbf{C}} \in \mathcal{C},\, \sum_{i,j}\hat{c}_{i,j} = k} \mathbb{E}\left[ F_\beta(\hat{\mathbf{C}}, \mathbf{C}) \right]
     = \argmax_{\hat{\mathbf{C}} \in \mathcal{C},\, \sum_{i,j}\hat c_{i,j} = k} \sum_{i,j} \hat c_{i,j}\mathbb{E}\left[ \frac{(1+\beta^2) c_{i,j}}{\beta^2 \sum_{i,j} c_{i,j} + k}\right].\label{eq:bayes_est_optim_inner}
\end{align}
The optimization expression from \eqref{eq:bayes_est_optim_inner} represents an inner maximization step, which is followed by an outer maximization step
\begin{align}
    \hat{\mathbf{C}}_{\text{Bayes}} &= \argmax_{\hat{\mathbf{C}} \in \{\hat{\mathbf{C}}_{\text{Bayes}}(0),\dots, \hat{\mathbf{C}}_{\text{Bayes}}(n_\mathcal{B})\}} \mathbb{E} \left[F_\beta(\hat{\mathbf{C}}, \mathbf{C}) \right].\label{eq:bayes_est_optim}
\end{align}

For any given $k$, the inner optimization problem in \eqref{eq:bayes_est_optim_inner} can be solved as a linear sum assignment problem (LSAP), subject to the constraint of $k$ links. In general, LSAP solvers can find unique (bipartite) pairings between elements of two sets
to maximize a user-specified total score. 
%
The constraint $\sum_{i,j}\hat c_{i,j} = k$ can be enforced using various methods, as discussed in \cite{ramshaw2012minimum}. Our innovation uses a data augmentation approach to incorporate the constraint into the original LSAP problem, as we now explain.

 
For each record pair $(\mathcal{A}_i, \mathcal{B}_j)$, let $\Delta_{i,j}^{(k)} = \mathbb{E}\left[(1+\beta^2)c_{i,j}/(\beta^2 \sum_{i,j} c_{i,j} + k)\right]$ represent a score for the pair; we  use these scores in the LSAP. Let $\boldsymbol{\Delta}^{(k)}$ be the $n_{\mathcal{A}} \times n_{\mathcal{B}}$ matrix with entry $\Delta_{i,j}^{(k)}$ in the $i$th row and $j$th column.
We create an augmented matrix $\tilde{\boldsymbol{\Delta}}^{(k)}$ of dimension $(n_\mathcal{A} + n_\mathcal{B} -k) \times n_{\mathcal{B}}$ with elements  in the $i$th row and $j$th column given by 
\begin{equation} \label{eq:augmented-matrix-weights}
    \tilde \Delta_{i,j}^{(k)} = 
    \begin{cases}
        \Delta_{i,j}^{(k)} & \text{if $i  \leq n_{\mathcal{A}}$,} \\
        M & \text{otherwise,}
    \end{cases}
\end{equation}
where 
\begin{equation}\label{eq:m}
    M = k (1+\beta^2)/\beta^2 \geq k \max_{i,j} \Delta_{i,j}^{(k)}.
\end{equation}
Letting $\tilde n_{\mathcal{A}} =n_\mathcal{A} + n_\mathcal{B} -k$, \eqref{eq:bayes_est_optim_inner} is transformed into the optimization problem,
\begin{align}\label{eq:lsap-augmented}
   & \argmax_{\mathbf{\hat C}}{\sum_{i=1}^{\tilde n_{\mathcal{A}}}\sum_{j=1}^{n_{\mathcal{B}}}\hat c_{ij}} \tilde \Delta^{(k)}_{ij} \\
    &\text{subject to,} \notag \\
    & \sum_{i=1}^{\tilde n_{{\mathcal{A}}}}\hat c_{i,j} = 1, j = 1, \dots, n_{\mathcal{B}}, \label{eq:lsap-augmented-c1} \\
    & \sum_{j=1}^{n_{\mathcal{B}}}\hat c_{i,j} =1, i = 1, \dots, \tilde n_{\mathcal{A}}.\label{eq:lsap-augmented-c2}
\end{align}
This is a variant of the general LSAP formulation.
The equality constraints in \eqref{eq:lsap-augmented-c1} and \eqref{eq:lsap-augmented-c2} ensure that all $n_\mathcal{B}$ elements of $\mathcal{B}$ have unique links to some elements represented by the rows $i \in \{1, \dots, n_{\mathcal{A}}, n_{\mathcal{A}}+1, \dots, n_{\mathcal{A}} + n_{\mathcal{B}}-k\}$ of $\tilde {\boldsymbol{\Delta}}^{(k)}$. 


The value of $M$ in \eqref{eq:m} is chosen to be large enough that the solution to \eqref{eq:lsap-augmented} links all elements in the augmented rows represented by $i \in \{n_{\mathcal{A}}+1, \dots, n_{\mathcal{A}} + n_{\mathcal{B}}-k\}$ to  $n_\mathcal{B} - k$ elements in $\mathcal{B}$, leaving only $k$ elements from the rows $i \in \{1, \dots, n_{\mathcal{A}}\}$ of $\mathcal{A}$ linked to $k$ elements of $\mathcal{B}$. Specifically, $M$ should be bounded below by $k$ times the largest element of $\Delta^{(k)}$. By setting $M$ this way, any reassignment of matches between $\mathcal{A}$ and $\mathcal{B}$ does not change the total score by more than $M$. As such, there will be exactly $n_\mathcal{B}-k$ links between elements of $\mathcal{B}$ and the ``dummy'' elements represented by $i \in \{n_{\mathcal{A}}+1, \dots, n_{\mathcal{A}} + n_{\mathcal{B}}-k\}$, and the remaining $k$ elements from $\mathcal{B}$ will be matched to elements of $\mathcal{A}$ represented by $i \in \{1, \dots ,n_{\mathcal{A}}\}$. By disregarding the $n_\mathcal{B}-k$ matches from the augmented rows, we isolate the top $k$ matches.  Per the objective function in \eqref{eq:lsap-augmented}, these $k$ matches maximize the score.

To illustrate how the solution to \eqref{eq:lsap-augmented} solves \eqref{eq:bayes_est_optim_inner}, consider a simple example where $n_\mathcal{A} = n_\mathcal{B} = 3$ and $k=2$. In this scenario, we construct the augmented $4 \times 3$ matrix $\tilde{\boldsymbol{\Delta}}^{(2)}$, as depicted in Figure \ref{fig:weight-matrix-ex}. Here, the $3 \times 3$ sub-matrix across the first three rows and columns represents $\boldsymbol{\Delta}^{(2)}$. The row labeled $\mathcal{A}_4$ represents the additional row, where each element is given the score $M = 4.0$. The value of $M$ is selected to exceed $k \max_{i,j} \Delta_{i,j}^{(2)} = 1.8$ per \eqref{eq:m}. The optimal solution to \eqref{eq:lsap-augmented} includes one link ($n_\mathcal{B}-k$ = 1) to the ``dummy'' row $\mathcal{A}_4$ and two links ($k=2$) to the records in $\mathcal{B}$. The augmented LSAP solver from \eqref{eq:lsap-augmented} guarantees that the resulting estimate $\hat{\boldsymbol{C}}^{(2)}$ (the circled record pairs in Figure \ref{fig:weight-matrix-ex}) achieves the highest score with exactly $k=2$ links.

\begin{figure}[t]
\centering
\begin{equation*}
\begin{array}{c|ccc}
 & \mathcal{B}_1 & \mathcal{B}_2 & \mathcal{B}_3 \\
\hline
\mathcal{A}_1 & 0.1 & 0.4 & \circled{0.9} \\
\mathcal{A}_2 & 0.2 & 0.5 & 0.8 \\
\mathcal{A}_3 & 0.3 & \circled{0.6} & 0.7 \\
\mathcal{A}_4 & \circled{4.0} & 4.0 & 4.0  \\
\end{array}
\end{equation*}
\caption{Example of an augmented matrix $\tilde{\boldsymbol{\Delta}}^{(2)}$ used in the LSAP.  The circled links optimize the total score for $k=2$ matches while respecting the bipartite matching.}
\label{fig:weight-matrix-ex}
\end{figure}

Since closed-form expressions for $\Delta_{i,j}^{(k)}$ are not available,
we use a Monte Carlo approximation based on samples from the posterior distribution of $\mathbf{C}$. Using the $L$ posterior samples  $\{\mathbf{C}^{(s)}: s=1, \dots, L\}$, we estimate each $\Delta_{i,j}^{(k)}$ using
\begin{equation}
    \Delta_{i,j}^{(k)} \approx \frac{1}{L} \sum_{s=1}^{L} \frac{(1+\beta^2)c_{i,j}^{(s)}}{\beta^2 \sum_{i,j} c_{i,j}^{(s)} + k}.
\end{equation}

These can be efficiently computed for all values of $(i, j)$ and $k$ by using appropriate data structures and exploiting the sparsity of $\mathbf{C}$. Following 
\cite{Sadinle2017}, suppose that samples from the posterior distribution of $\mathbf{C}$ are represented as $n_{\mathcal{B}} \times 1$ vectors $\mathbf{Z}^{(s)} = (Z^{(s)}_1, \dots, Z^{(s)}_{n_{\mathcal{B}}})$ where any $Z^{(s)}_j = i$ when $c_{i,j}^{(s)}=1$ and $Z^{(s)}_j = n_{\mathcal{A}}+j$ when $c_{i,j}^{(s)}=0$. Let $\tilde{ \mathbf{Z}}$ be the $ n_{\mathcal{B}} \times L$ matrix comprising the columns $[\mathbf{Z}^{(1)}, \dots, \mathbf{Z}^{(L)}]$; and, let $\mathbf{I} = [I_{j}^{(s)}]$ be the $n_{\mathcal{B}} \times L$ matrix with elements  $I_{j}^{(s)} = 1$ when $Z^{(s)}_j \leq n_{\mathcal{A}}$, i.e., when record $j\in \mathcal{B}$ for iteration $s$ matches an element of $\mathcal{A}$, and $I_{j}^{(s)} = 0$ otherwise. 
From $\tilde{\mathbf{Z}}$, we reconstruct the $n_{\mathcal{A}} n_{\mathcal{B}} \times L$ sparse matrix $\tilde{\mathbf{C}} = [c_{i,j}^{(s)}]$ with $n_{\mathcal{A}} n_{\mathcal{B}}$ rows corresponding to pairs $(\mathcal{A}_i, \mathcal{B}_j)$ and columns corresponding to posterior samples $s\in \{1,\dots, L\}$. For computational purposes, this matrix is represented internally in coordinate list format \citep{Bates2023} and computed with time complexity $\mathcal{O}(n_{\mathcal{B}}L)$ by iterating through all elements of $\tilde{\mathbf{Z}}$. Then, for a given $k$, the matrix $\boldsymbol{\Delta}^{(k)}$ can be vectorized by computing
\begin{equation}\label{eq:weights_computation}
   Vec(\boldsymbol{\Delta}^{(k)}) \approx {\tilde{\mathbf{C}}} \frac{(1+\beta^2)/L}{\beta^2 {\bf I}^T{\bf 1} + k},
\end{equation}
where matrix division is understood to be element-wise. Using the sparse matrix representation of $\tilde{\mathbf{C}}$, the multiplication between ${\tilde{\mathbf{C}}}$ and the column vector $((1+\beta^2)/L)/(\beta^2 {\bf I}^T{\bf 1} + k)$ in \eqref{eq:weights_computation} is computed with time complexity $\mathcal{O}(n_{\mathcal{B}}L)$.

\subsubsection{An Exact Solution to $\hat{\mathbf{C}}_{\text{OS}}$}\label{sec:exact}

To compute $\hat{\mathbf{C}}_{\text{OS}}$ in \eqref{eq:c-ml}, we again use the  framework of \citet{jansche-2007-maximum} and first find the inner maximization solutions $\hat{\mathbf{C}}_{\text{OS}}(k)$ satisfying
\begin{align}
     \hat{\mathbf{C}}_{\text{OS}}(k) = \argmax_{\hat{\mathbf{C}} \in \mathcal{C},\, \sum_{i,j}\hat c_{i,j} = k} \sum_{i,j} \hat c_{i,j} \frac{(1+\beta^2) \hat p_{i,j}}{\beta^2 \sum_{i,j} \hat p_{i,j} + k}.\label{eq:os_est_optim_inner}
\end{align}
We then use these estimates to determine the outer maximization solution from
\begin{align}
    \hat{\mathbf{C}}_{\text{OS}} &= \argmax_{\hat{\mathbf{C}} \in \{\hat{\mathbf{C}}_{\text{OS}}(0),\dots, \hat{\mathbf{C}}_{\text{OS}}(n_\mathcal{B})\}}F_\beta(\hat{\mathbf{C}}, \hat{\mathbf{P}}),\label{eq:os_est_optim}
\end{align}
where $\hat{\mathbf{P}}=[\hat p_{ij}]$ is the matrix containing all $\hat{p}_{ij}$.

The algorithm follows the same steps as the one used to approximate $\hat{\mathbf{C}}_{\text{Bayes}}$, except we replace $\Delta_{i,j}^{(k)}$ with its plug-in estimate, $\hat \Delta_{i,j}^{(k)} = ((1+\beta^2) \hat{p}_{i,j})/(\beta^2 \sum_{i,j}\hat {p}_{i,j} + k)$. Since no Monte Carlo approximations are used, the algorithm produces an exact solution for \eqref{eq:c-ml}. 



\section{Estimation of Overlap Population Size}\label{sec:estimation-overlap}

An important characteristic of any linkage structure $\mathbf{C}$ is the induced number of matching records, $\sum_{i,j} c_{i,j}$.  We refer to this quantity as the overlap population size. 
The overlap size is directly related to the size of the joint population after removing duplicates, which we write as  $n_{\mathcal{A}, \mathcal{B}} = n_{\mathcal{A}} + n_{\mathcal{B}} - \sum_{i,j} c_{i,j}$.  Given an estimate $\hat{\mathbf{C}}$, we  define the estimated overlap size as $\sum_{i,j} \hat{c}_{i,j}$; similarly, we can define the estimated population size as $\hat n_{\mathcal{A}, \mathcal{B}} = n_{\mathcal{A}} + n_{\mathcal{B}} - \sum_{i,j} \hat c_{i,j}$. 

In some record linkage applications, one objective is to estimate the overlap population size accurately. 
For example, \citet{wortmanthesis} links a file comprising voters in North Carolina who cast provisional ballots to the official state voter registration file.  The overlap size represents the number of voters on the official voter registration file---which includes current and no-longer registered voters---who cast provisional ballots.  This quantity is needed to study the effects of local policies that removed voters from the registration rolls.  As another example, in checking the quality of the decennial census, the Census Bureau matches records from the collected decennial population census data to records from a post-enumeration survey. They need to estimate the number of individuals who appear in both data sources, i.e., the overlap size, for use in assessments of undercount and overcount estimation (\url{https://www.census.gov/programs-surveys/decennial-census/about/coverage-measurement/pes.html}).

Since the $F$-score equally balances precision and recall (when choosing $\beta =1$), we can expect the $F$-score optimal linkage to produce accurate estimates of overlap population size. To motivate this, note that the precision $P$ is defined as ratio of the number of true positive links, $\sum_{i, j} c_{i,j} \hat c_{i,j}$, to the true overlap size, $\sum_{i,j} c_{i,j}$. Similarly, the recall $R$ is defined as the number of true positive links divided by the estimated overlap size $\sum_{i,j} \hat c_{i,j}$. Consequently, when $P=R$ for a specific $\hat{\mathbf{C}}$, the induced overlap size estimate should equal the true overlap size, i.e., $\sum_{i,j}\hat c_{i,j} = \sum_{i,j}c_{i,j}$. This simple property suggests choosing an equal balance between precision and recall in the definition of the $F$-score to be optimized. 






Of course, we must evaluate the potential accuracy of the estimated overlap population size using the $F$-score approximations from Section \ref{sec:F-score-algo}.  Before doing so, however, we provide additional motivation for considering this new point estimator as compared to existing methods.  Here, we focus on the Bayes estimator introduced by \cite{Sadinle2017}. While we do not delve into other methods,
we highlight that these methods generally do not  directly optimize for accurate overlap size and thus may run into similar issues. 

\subsection{Definition of the BRL Estimator} 
\label{sec:def_brl_estimator}

We begin by defining the Bayes estimator of \citet{Sadinle2017}, which we call the Beta Record Linkage (BRL) estimator.
Using the definition of $\mathbf{Z}$ presented in Section \ref{sec:approx}, let  $\hat{\mathbf{Z}} = (\hat Z_1,  \dots,\hat  Z_{n_{\mathcal{B}}})$ be the vector-representation of the estimated linkage structure $\hat{\mathbf{C}}$.  
\citet{Sadinle2017} introduced a Bayes estimator for $\mathbf{Z}$ based on an additive loss  $L(\mathbf{Z}, \hat{\mathbf{Z}})$ parameterized by positive constants $(\lambda_{10}, \lambda_{01}, \lambda_{11}).$ We have 
\begin{equation} \label{sadinle-additive-loss}
    L(\mathbf{Z}, \hat{\mathbf{Z}}) = \sum_{j=1}^{n_{\mathcal{B}}}L_j(Z_j, \hat{Z}_j),
\end{equation}
where
\begin{align*}
L_j(Z_j, \hat{Z_j})=
    \begin{cases}
        0 & \text{if $Z_j \leq n_{\mathcal{A}}$}, \\
        \lambda_{10} & \text{if $Z_j \leq n_{\mathcal{A}}$, $\hat{Z}_j = n_{\mathcal{A}} + j$}, \\
        \lambda_{01} & \text{if $Z_j = n_{\mathcal{A}} +j$, $\hat{Z}_j  \leq n_{\mathcal{A}}$},  \\
        \lambda_{11} & \text{if $Z_j, \hat{Z}_j \leq n_{\mathcal{A}}$, $\hat{Z}_j \neq Z_j$}.
    \end{cases}
\end{align*}
Here,  $\lambda_{10}$ is the loss incurred for any record $\mathcal{B}_j$ from deciding it has no link among $\mathcal{A}$ when in fact it does; $\lambda_{01}$ is the loss incurred for any record $\mathcal{B}_j$ from deciding it links to some record in $\mathcal{A}$ when in fact its match is not in $\mathcal{A}$; and, $\lambda_{11}$ is the loss incurred for any record $\mathcal{B}_j$ from deciding it is linked to some record $\mathcal{A}_i$ when in fact its match is some other record $\mathcal{A}_{k}$, where $k \leq \mathcal{A}$.

The BRL estimator of \citet{Sadinle2017} is 
\begin{equation}
    \hat{\mathbf{Z}}_{\text{BRL}} = \argmin_{\hat{\mathbf{Z}}} \mathbb{E}[L(\mathbf{Z}, \hat{\mathbf{Z}})].\label{eq:bayesMS}
\end{equation}

Suppose that we have the (approximate) posterior distribution of $\mathbf{Z}$ given the data $\gamma$ used to match records across $\mathcal{A}$ and $\mathcal{B}$; we write this as $p(\mathbf{Z}| \gamma)$.
We give an example of $\gamma$ in Section \ref{sec:brl-primer}.
The BRL estimator can be computed by solving a standard LSAP.
However, in this context, we find the minimizer and use a matrix for the optimization step that is interpreted in terms of costs rather than scores. In this context, the $(i,j)$ element of the $(n_\mathcal{A}+n_\mathcal{B}) \times n_\mathcal{B}$ cost matrix is  
\begin{equation}\label{sadinle-lsap}
    w_{ij}  = 
    \begin{cases}
        \lambda_{01}P(Z_j = n_{\mathcal{A}}+j|\gamma) + \lambda_{11}P(Z_j \notin \{i, n_{\mathcal{A}}+j\}|\gamma) & \text{if $i  \leq n_{\mathcal{A}}$} \\
        \lambda_{10}P(Z_j \neq n_{\mathcal{A}}+j|\gamma) & \text{if $i = n_{\mathcal{A}} + j$} \\
        \infty & \text{otherwise.}
    \end{cases}
\end{equation}

\citet{Sadinle2017} shows that \eqref{eq:bayesMS} has a closed-form solution when both $ 0 < \lambda_{10} \leq \lambda_{01}$ and $\lambda_{11} \geq \lambda_{10} + \lambda_{01}$. Specifically, for $j=1, \dots, n_{\mathcal{B}}$, he shows that  
\begin{equation} \label{sadinle-closed-form}
\hat{Z}_j = 
    \begin{cases}
        i & \text{if $P(Z_j=i|\gamma) > \frac{\lambda_{01}}{\lambda_{01}+\lambda_{10}} + \frac{\lambda_{11} - \lambda_{01}-\lambda_{10}}{\lambda_{01}+\lambda_{10}}P(Z_j \notin \{i, n_{\mathcal{A}}+j\}|\gamma)$} \\
        n_{\mathcal{A}}+j & \text{otherwise.} 
    \end{cases}
\end{equation}
In this case, a necessary condition for linking the pair $(\mathcal{A}_i, \mathcal{B}_j)$ is that  $P(Z_j=i|\gamma) > 0.5$ \citep{Sadinle2017}. This  is apparent when adopting the default parameters suggested by \citet{Sadinle2017}, namely $ \lambda_{10} = \lambda_{01} = 1$ and $\lambda_{11} = 2$. These default parameter settings are used in both the simulation and analyses of Section \ref{sec:sim-test}. 

\subsection{Conservative Nature of the BRL Estimator} \label{sec:brl_calibration_challenge}

Using the estimator in \eqref{sadinle-closed-form}, \citet{Sadinle2017} observes that the LSAP algorithm adopts a conservative approach in declaring links, in that it requires $P(Z_j=i|\gamma) \geq .50$.
This reduces the risk of incorrect linkages, but it can introduce bias in the estimation of overlap size under certain linkage scenarios. As an example, consider a task of linking unique individuals in a large data file  $\mathcal{A}$---which has near complete coverage of a population---to participants in a survey data file $\mathcal{B}$ that overlaps with  $\mathcal{A}$.  Within $\mathcal{A}$, suppose exactly two distinct individuals $\mathcal{A}_{1}$ and $\mathcal{A}_{2}$ share identical names with individual $\mathcal{B}_1 \in \mathcal{B}$. As a result, the posterior probabilities that $Z_1 = 1$ or $Z_j=2$ are both near (but necessarily below) $0.5$, as exemplified in  Table \ref{tab:ex1-con}. 
In this case, it is reasonable to conclude that one of  $\mathcal{A}_{1}$ or $\mathcal{A}_{2}$ is the true link to $\mathcal{B}_1$, and therefore to count $\mathcal{B}_1$ as part of the overlap population. However, if we use \eqref{sadinle-closed-form}, the BRL estimate fails to declare any links for $\mathcal{B}_1$ in $\mathcal{A}$, even though the chance that  
$\mathcal{B}_1$ is not linked to any elements in $\mathcal{A}$ is only 2\%. Similar issues can arise when more than two records in $\mathcal{A}$ are highly plausible links for some $\mathcal{B}_j$.


\begin{table}[t]
\centering
\centering
\caption{Example of a linkage scenario where BRL selects no match for $\mathcal{B}_1$ when arguably $\mathcal{B}_1$ should be counted as part of the overlap population.}
\label{tab:ex1-con}
\begin{tabular}{|l|l|}
\hline
\rule[-1ex]{0pt}{3.5ex}  Index $i$ for $\mathcal{A}_i$ or for non-match & $P(Z_{1}=i| \gamma)$  \\
\hline\hline
\rule[-1ex]{0pt}{3.5ex}  $1$ & $0.49$ \\
\hline
\rule[-1ex]{0pt}{3.5ex}  $2$ & $0.49$   \\
\hline
\rule[-1ex]{0pt}{3.5ex}  $3$ & $0.00$  \\
\hline
\rule[-1ex]{0pt}{3.5ex}  $4$ & $0.00$  \\
\hline
\rule[-1ex]{0pt}{3.5ex}  $5$ & $0.00$  \\
\hline
\rule[-1ex]{0pt}{3.5ex}  $n_{\mathcal{A}}+1$ (non-match) & $0.02$  \\
\hline
\end{tabular}
\end{table}

\begin{table}[t]
\centering
\caption{Example of a general scenario where the LSAP algorithm does not declare a link for $\mathcal{B}_j$ when arguably it should be counted as part of the overlap population.}
\label{mod}
\begin{tabular}{|l|l|} 
\hline
\rule[-1ex]{0pt}{3.5ex}  Index $i$ for $\mathcal{A}_i$ or for non-match & {$P(Z_{j}=i| \gamma)$}  \\
\hline\hline
\rule[-1ex]{0pt}{3.5ex}  $1$ & $0.25$ \\
\hline
\rule[-1ex]{0pt}{3.5ex}  $2$ & $0.25$   \\
\hline
\rule[-1ex]{0pt}{3.5ex}  $3$ & $0.10$  \\
\hline
\rule[-1ex]{0pt}{3.5ex}  $4$ & $0.09$  \\
\hline
\rule[-1ex]{0pt}{3.5ex}  $5$ & $0.01$  \\
\hline
\rule[-1ex]{0pt}{3.5ex}  $n_{\mathcal{A}}+j$ (non-match) & $0.30$  \\
\hline
\end{tabular}
\end{table}


It is important to note that \eqref{sadinle-lsap} can produce conservative estimates of links even when the conditions in \eqref{sadinle-closed-form} do not apply. This is evident in the following proposition.

\begin{proposition}[Sufficient condition for a non-match] \label{suff1}
The BRL estimator declares record $\mathcal{B}_j$ a non-link, i.e., $Z_j =n_{\mathcal{A}} + j$, if
\begin{equation} \label{eq:suff1}
     (\lambda_{10} + \lambda_{01} - \lambda_{11})P(Z_j=n_{\mathcal{A}}+j|\gamma) \geq \lambda_{10} - \lambda_{11}  + \lambda_{11}\max_{i: i\leq n_\mathcal{A}}P(Z_j=i|\gamma).
\end{equation}
\begin{proof}
Suppose that the bipartite matching condition in \eqref{suff1} holds. Then, by assumption in the proposition, for all $i \in \{1,  \dots, n_{\mathcal{A}}\}$ we have 
\begin{align}
    &(\lambda_{10} + \lambda_{01} - \lambda_{11}) P(Z_j=n_{\mathcal{A}}+j) \geq \lambda_{10} - \lambda_{11} + \lambda_{11}P(Z_j=i|\gamma).
\end{align}    
This implies that 
\begin{align}
    &\lambda_{01} P(Z_j=n_{\mathcal{A}}+j) + \lambda_{11}(1-P(Z_j =i|\gamma) - P(Z_j= n_{\mathcal{A}}+j|\gamma)) \geq \lambda_{10}(1-P(Z_j = n_{\mathcal{A}} +j|\gamma)).
    \end{align}
As a result, we have 
\begin{align}
    &\lambda_{01} P(Z_j=n_{\mathcal{A}}+j|\gamma) + \lambda_{11}P(Z_j \notin \{i, n_{\mathcal{A}}+j\}|\gamma) \geq \lambda_{10}P(Z_j\neq  n_{\mathcal{A}} +j|\gamma).
\end{align}
The cost matrix defined by \eqref{sadinle-lsap} implies that the LSAP algorithm will choose $Z_j = n_\mathcal{A}+j$ for record $\mathcal{B}_j$. Note that the converse does not generally hold, as there can be instances where the LSAP solver opts for $Z_j = n_\mathcal{A}+j$, yet the condition in \eqref{eq:suff1} is not satisfied. 
\end{proof}
\end{proposition}

For any individual $\mathcal{B}_j$, Proposition \ref{suff1} shows the connection between the posterior probabilities of having no links and of the most likely link candidate in $\mathcal{A}$. That is, given pre-chosen cost parameters $(\lambda_{10}, \lambda_{01}, \lambda_{11})$, with a sufficiently low maximum link probability, i.e., $\max_{i: i\leq n_\mathcal{A}}P(Z_j=i|\gamma)$, the LSAP algorithm always declares a non-link for record $j$. 

For different choices of fixed cost parameters, we obtain different variations of the sufficient condition. A particularly illuminating result follows under unity cost ($\lambda_{10} = \lambda_{01} = \lambda_{11}$), which we state as a corollary below. 
\begin{corollary}[Sufficient Condition for non-match under unity costs]
    Under the unity cost assumption ($\lambda_{10} = \lambda_{01} = \lambda_{11}$), a sufficient condition for a non-match for record $\mathcal{B}_j$ is     
\begin{equation}\label{suff2_eq}
        P(Z_j=n_{\mathcal{A}}+j) \geq \max_{i: i\leq n_\mathcal{A}}P(Z_j = i).
\end{equation}    
\end{corollary}
The sufficient condition in \eqref{suff2_eq} assigns a non-link to $\mathcal{B}_j$ when its posterior probability of having no links is largest.
This particular decision rule is reasonable when the posterior distribution for $Z_j$ is characterized by large probability mass either at $Z_j=n_{\mathcal{A}} + j$ or $Z_j = i$ for some $i$. However, if the posterior distribution is multi-modal  across a variety of potential links in $\mathcal{A}$, the sufficient condition may be undesirable.  To illustrate, consider the  example in Table \ref{mod}. 
According to \eqref{suff2_eq}, the LSAP will result in a non-link decision. This results mainly because of significant uncertainty in the match status, as reflected through a posterior distribution with moderate probability masses placed on the decisions $Z_j=1$, $Z_j=2$, and $Z_j=n_{\mathcal{A}}+j$. In a comparison space with multiple instances like the one in Table \ref{mod}, all would be designated non-links by the LSAP. Such behavior may lead to underestimation of overlap size estimation in practice; for example, in Table \ref{mod}, the posterior probability of there being a match, $1-P(Z_j=n_\mathcal{A}+j|\gamma)$, is quite high at $70\%$.

\section{Simulation Studies and Illustrative Examples}
\label{sec:sim-test}

We now evaluate the point estimator based on the $F$-score using simulation studies and  applications. As a comparison, we also evaluate the BRL estimator described in Section \ref{sec:def_brl_estimator}. To obtain both, we estimate the posterior distribution of $\mathbf{Z}$ using the Bayesian record linkage model of \cite{Sadinle2017}, 
which we summarize in Section \ref{sec:brl-primer}.
In section \ref{sec:simulation}, we consider a simulation study with varying quality of the variables used for matching and varying size of the overlap, i.e., the number of records in $\mathcal{A}$ and $\mathcal{B}$ where $\mathcal{A}_{i0}=\mathcal{B}_{j0}$. In Section \ref{sec:real-data}, we consider genuine record linkage applications for which ground truth is available.

We investigate three  questions.  First, does using the $F$-score estimator result in higher quality record linkage performance?  We evaluate this by computing the $F$-scores for the point estimates using ground truth data.  Second, does using the $F$-score estimator provide good overlap population size estimation? We evaluate this by computing the overlap size induced from the linkage estimate and comparing it to true overlap.  Third, is using the $F$-score estimator compatible with model predictions, particularly credible intervals for the population size obtained from the Bayesian record linkage model? We evaluate this by assessing whether the estimated overlap is inside the $95\%$ credible interval.


\subsection{Bayesian Bipartite Record Linkage Model}
\label{sec:brl-primer}

Suppose each record $\mathcal{A}_i \in \mathcal{A}$ and $\mathcal{B}_j \in \mathcal{B}$ has $F$ shared attributes, which we call linking fields. For any $i$ and $j$, let $\mathcal{A}_i = (\mathcal{A}_{i,1}, \dots, \mathcal{A}_{i,F})$ and $\mathcal{B}_j = (\mathcal{B}_{j,1}, \dots, \mathcal{B}_{j,F})$, and 
let $\gamma_{i,j}=(\gamma_{i,j}^{(1)}, \dots, \gamma_{i,j}^{(F)})$ be the   comparison vector derived from the linking fields for the record pair. 
Each $\gamma_{i,j}^{(f)}$ is the result of the comparison between $\mathcal{A}_{i,f}$ and $\mathcal{B}_{j,f}$.  For instance, $\gamma_{i,j}^{(f)}$ can be a binary indicator of whether field $f$ is identical for $\mathcal{A}_i$ and $\mathcal{B}_j$. 
We assume that each $\gamma_{i,j}^{(f)}$ takes on $d_f \geq 2$ possible  agreement levels. 

We suppose that each $\gamma^{(f)}_{i,j}$ is a realization of a random variable $\Gamma^{(f)}_{i,j}$.  Let $\Gamma_{i,j}=(\Gamma_{i,j}^{(1)}, \dots, \Gamma_{i,j}^{(F)})$. 
For $l = 1, \dots, d_f$, let $m_{f,l} = P(\Gamma_{ij}^{(f)}=l | Z_j = i)$ and $u_{f,l} = P(\Gamma_{ij}^{(f)}=l | Z_j \neq i)$ denote the probability that field $f$ takes on value $l$ for matches and non-matches, respectively. The model of \cite{Sadinle2017} 
assumes the $\Gamma_{i,j}^{(f)}$ are conditionally independent given $Z_j$, so that  
\begin{eqnarray}
    &  P(\Gamma_{i,j}=\gamma_{i,j} \mid Z_{j}=i) = \prod_{f=1}^{F}\prod_{l=1}^{d_f} m_{f,l}^{\mathbbm{I}(\gamma_{i,j}^{(f)}=l)}\label{eq:simulate gamma_link}\\
    &  P(\Gamma_{ij}=\gamma_{ij} \mid Z_{j}\neq i)  =\prod_{f=1}^{F}\prod_{l=1}^{d_f} u_{f,l}^{\mathbbm{I}(\gamma_{i,j}^{(f)}=l)}.\label{eq:simulate gamma_non-link}
\end{eqnarray}
\cite{Sadinle2017} uses (uniform) Dirichlet prior distributions for each $\mathbf{m}_f = (m_{f,1}, \dots, m_{f,d_f})$ and $\mathbf{u}_f = (u_{f,1}, \dots, u_{f,d_f})$, and a prior distribution on $\mathbf{Z}$ that enforces the bipartite matching constraint. 

The model parameters can be estimated using a Markov chain Monte Carlo sampler.  This results in $L$ plausible draws of the linkage structure as represented by $(\mathbf{Z}^{(1)}, \dots, \mathbf{Z}^{(L)})$.  In the simulations, we use the ``BRL'' package \citep{BRL_package} available in the software R to estimate the posterior distribution and obtain these draws. 


\subsection{Simulation Study}
\label{sec:simulation}




We consider several data scenarios characterized by different overlap sizes and error levels.  To facilitate repeated sampling computations, we let $n_{\mathcal{A}}= 1000$ and $n_{\mathcal{B}} = 50$.  We determine 
the overlap size by setting the proportion $\pi$ of records in $\mathcal{B}$ that have links in $\mathcal{A}$ as $\pi  \in \{25\%, 50\%, 75\%, 100\%\}$. In each scenario, we generate comparison vectors for $F=3$ binary fields, where $\gamma_{i,j}^{(f)} = 1$ when $\mathcal{A}_{i,f}=\mathcal{B}_{j,f}$  and $\gamma_{i,j}^{(f)} = 2$ otherwise. For each field $f$, we set $\mathbf{m}_f = (m_{f,1}, m_{f,2})$
and $\mathbf{u}_f =(u_{f,1}, u_{f,2})$ 
to represent one of three error levels, where smaller values in ${m}_{f,1}$ accompanied by smaller values in ${u}_{f,2}$ indicate increased error levels in the linking fields. The parameter settings for each  level are displayed in Table \ref{tab:sim:mandu}.  These parameter choices are guided by the values of $\mathbf{m}_f$ and $\mathbf{u}_f$ from the RLdata500 data  \citep{RecordLinkage} that we use in section \ref{sec:rldata}.  Briefly,  the low error scenario represents minimal errors across all fields. The moderate and moderate-high error settings represent situations where errors are introduced into two of the three fields to varying extents. When referring to simulation results from the $F$-score point estimator and BRL estimator, we use F-Algo and BRL, respectively.

\begin{table}[t]
\centering
\caption{The $\mathbf{m}$ and $\mathbf{u}$ parameters for the three simulation scenarios. Here, we present the probabilities that the fields match for linked records ($m_{f,1}$) and do not match for non-linked records ($u_{f,2}$).}
\label{tab:sim:mandu}
\begin{tabular}{lllllll}
\hline
Error Level & $m_{1,1}$ & $m_{2,1}$ & ${m}_{3,1}$ & ${u}_{1,2}$ & ${u}_{2,2}$ & ${u}_{3,2}$ \\\hline
Low  & 0.93 & 0.93 & 0.98 & 0.94 &  0.94 & 0.98\\
Moderate & 0.83 & 0.83 & 0.98 & 0.84 & 0.84 & 0.98\\
Moderate-High & 0.83 & 0.83 & 0.88 & 0.84 & 0.84 & 0.98\\
\hline
\end{tabular}
\end{table}

\begin{table}[t]
  \centering
\caption{Average $F$-score, estimated file overlap, true overlap, and model-based overlap 95\% credible interval for $N=1000$ replicates in the simulation study. $F$-score and estimated overlap are computed from the BRL and F-Algo point linkage estimates.}
  \label{tab:sim-result-real}
  \begin{tabular}{lccccccr}
    \toprule
    \multirow{2}{*}{Error Level} & \multirow{2}{*}{Overlap} & 
    \multicolumn{2}{c}{$F$-score} & \multicolumn{2}{c}{Est. Overlap} & 
    \multicolumn{1}{c}{True Overlap} & \multicolumn{1}{c}{Overlap} \\
    \cmidrule{3-6}
    & & BRL & F-Algo & BRL & F-Algo & 
    \multicolumn{1}{c}{} & \multicolumn{1}{c}{95\% CI} \\
    \midrule
    \multirow{4}{*}{Low} & 25\% & .84 & .88 & 13 & 14 & 
    13 & (14,50) \\
    & 50\% & .89 & .91 & 27 & 28 & 
    24 & (27,49) \\
    & 75\% & .88 & .86 & 33 & 35 & 
    37 & (36,50) \\
    & 100\% & .95 & .95 & 46 & 46 & 
    50 & (46,50) \\
    \midrule
    \multirow{4}{*}{Moderate} & 25\% & 0 & .44 & 0 & 23 & 
    13 & (0,48) \\
    & 50\% & 0 & .45 & 0 & 28 & 
    24 & (2,49) \\
    & 75\% & 0 & .52 & 0 & 32 & 
    37 & (11,50) \\
    & 100\% & .60 & .59 & 29 & 41 & 
    50 & (34,50) \\
    \midrule
     \multirow{4}{*}{Moderate-high} & 25\% & 0 & .24 & 0 & 20 & 
    13 & (0,48) \\
    & 50\% & 0 & .39 & 0 & 30 & 
    27 & (0,48) \\
    & 75\% & 0 & .50 & 0 & 31 & 
    37 & (8,50) \\
    & 100\% & .48 & .50 & 25 & 38 & 
    50 & (27,50) \\

    \bottomrule
  \end{tabular}
\end{table}

Table \ref{tab:sim-result-real} displays the simulation results. For each metric, the averages are computed across $N=1000$ independent simulation runs to ensure the Monte-Carlo errors remain below $10^{-6}$. At low error levels, the BRL estimator and the F-Algo estimator have comparable efficacy, both in the $F$-scores and overlap size estimates. In scenarios with moderate to moderate-high noise, the F-Algo estimator consistently outperforms the BRL estimator in the sense of higher $F$-scores and more accurate overlap size estimates.  In these settings, the BRL estimator is overly conservative. In fact, it often declares all potential pairs as non-links, leading to  $F$-scores of zero.  Not surprisingly, the absolute performance of both estimators declines as overlap size decreases. With the lowest overlap and at least moderate errors, F-Algo tends to assign too many record pairs as links, whereas BRL tends to declare all record pairs as non-links.  


\subsection{Illustrative Examples}\label{sec:real-data}
In this section, we evaluate the F-Algo estimator using data where the ground truth is established. We begin by validating its performance on the  RLdata500  \citep{RecordLinkage}.  RLdata500 is frequently used in the record linkage literature due to its simplicity and low amount of error  \citep{steorts2015entity}. We then test its effectiveness using the Union Army data \citep{fogel2000aging}.

\subsubsection{Linkage with the RLdata500} \label{sec:rldata}


RLdata500 is a synthetic dataset comprising 500 personal information records, 50 of which are noisy duplicates. At most two records refer to the same person. The linking fields include components of names and birth date.  

We construct a bipartite version of these data as follows. First, we split non-duplicated records at random between $\mathcal{A}$ and $\mathcal{B}$. Second, for duplicated records, we place the first record instance in  $\mathcal{A}$ and the second instance in  $\mathcal{B}$.  For string attributes (first and last name), we construct $\gamma_{i,j}^f$ using  normalized Levenshtein distance thresholds at $(0, 0.25, 0.5, 1)$, which is the default in the ``BRL'' package  \citep{BRL_package}. For numeric attributes (birth year, birth month, and birth day), we use binary $\gamma_{i,j}^f$ to indicate exact match on each field.

We consider four versions of the Bayesian record linkage model of section \ref{sec:brl-primer}.
The four models use different attributes  for linkage. Model A only uses birth year, birth month, and birth day.  Model B only uses the last name and birth year.  Model C uses first name, last name, and birth year.  Model D uses first name, last name, birth year, birth month, and birth day. We fit the models using the ``BRL'' package with default hyperparameters, collecting 20,000 posterior samples after a burn-in of 5,000 iterations. 

\begin{table}[t]
    \centering
\begin{tabular}{@{}rcccccc@{}}
\toprule
                 & \multicolumn{2}{c}{$F$-score} & \multicolumn{2}{c}{Est. Overlap} & True Overlap & Overlap \\ \cmidrule(r){2-5}
                 & {BRL}          & {F-Algo}         & {BRL}            & {F-Algo}           &                & 95\% CI                  \\ \midrule
 {Model A} & 0.71         & 0.71                    & 32             & 32               & 50             & (40, 122)         \\

{Model B} & 0.76         & 0.78                    & 47             & 52               & 50             & (37, 109)         \\
{Model C} & 0.90         & 0.90                 & 43             & 43               & 50             & (43, 58)          \\
{Model D} & 0.98         & 0.98                 & 51             & 51               & 50             & (49, 54)          \\ \bottomrule
\end{tabular}\caption{$F$-score, estimated file overlap, true overlap, and model-based overlap 95\% credible interval for the four models estimated using RLdata500. $F$-score and estimated overlap are computed from the BRL and F-Algo point linkage estimates.}
    \label{tab:rldatabipartite}
\end{table}

Table \ref{tab:rldatabipartite} displays the results for the RLdata500 illustration. 
For Model A, Model C, and Model D, the $F$-score and estimated overlap size for F-Algo and BRL are identical. For model B, the F-Algo estimate has a marginally higher $F$-score than the BRL estimate. In all cases, the estimated overlap size is inside the 95\% credible interval. These observations are line with results from the simulation study, where the two estimators offer similar results in the absence of substantial errors in the linking fields. 
These results show that directly maximizing the expected $F$-score is compatible with anticipated behavior in these benchmark data.

\subsubsection{Linkage with the Union Army Data} \label{sec:Union}

The Union Army data comprise a longitudinal sample of Civil War veterans collected as part of the Early Indicators of Aging project \citep{fogel2000aging}. Records of soldiers from 331 Union companies were collected and carefully linked to a data file comprising military service records---which we call the MSR file---as well as other sources. 
These records also were  linked to the 1850, 1860, 1900, and 1910 censuses. 
The quality of the linkages in this project is considered very high, as the true matches were manually made by experts \citep{fogel2000aging}.  Thus, the Union Army data file can be used to test automated record linkage algorithms.

We consider re-linking soldiers from the {MSR} data to records from the 1900 census, which we call the {CEN} data file. This linkage problem is difficult for automated record linkage algorithms due to the presence of soldiers' family members in the {CEN} data. Furthermore, not all soldiers from the {MSR} data have a match in  the {CEN} data. However, we can consider the linkages identified by the Early Indicators of Aging project as truth.  For the linking fields, we use first name, last name, middle initial, and approximate birth year.  

We use two types of blocking to reduce the number of comparisons: on birth place, or on last name initial.  For each scheme, we estimate the Bayesian record linkage model separately in the blocks.  The first blocking scheme uses birth place.  Here, we present results for a block comprising all records with birth place of Michigan; we call this Block 1.
This block has 529 records in the MSR data and 1840 records in the CEN data.
The second blocking scheme uses the first letter of the last name. 
Here, we present results for a block comprising all records with last name starting with ``O'' as in Osborn or O'Connell; we call this Block 2. This block has 504 records in the MSR data and 599 records in the CEN data. 
We note that blocking based on last name initial rather than birth place tends to result in more erroneous linkages.
Individuals are more likely to have the same or a similar last name if they are in grouped by last name initial than if they are grouped by birth place.

We consider two versions of comparison vectors, which results in two record linkage models. Model A uses Levenshtein distances to compare names with thresholds $(0, 0.25, 0.5, 1)$, a binary comparison for middle initial, and a three-level comparison with threshold for birth year. Here, we quantize differences in birth years to the bins $[0, 1]$, $(1, 5]$, and $(5, \infty)$. Model B is a slight modification, using the thresholds $(0, 0.1, 0.5, 1)$ for Levenshtein comparisons instead.

\begin{table}[t]
\centering
\begin{tabular}{@{}rrcccccr@{}}
\toprule
\multicolumn{2}{l}{\multirow{2}{*}{}}               & \multicolumn{2}{c}{$F$-score} & \multicolumn{2}{c}{Est. Overlap} & True Overlap & Overlap\\ \cmidrule(lr){3-6}
\multicolumn{2}{l}{}                                & {BRL}          & {F-Algo}        & {BRL}            & {F-Algo}           &              & 95\% CI                \\ \midrule
\multirow{2}{*}{Block 1} & {Model A} & 0.87         & 0.87          & 154            & 161              & 188          & (150, 178)      \\
                                  & {Model B} & 0.74         & 0.75          & 117            & 138              & 188          & (125, 161)      \\ \midrule
\multirow{2}{*}{{Block 2}} & {Model A} & 0.51         & 0.66          & 58             & 88               & 144          & (77, 110)       \\
                                  & {Model B} & 0.23         & 0.57          & 22             & 88               & 144          & (59, 99)        \\ \bottomrule
\end{tabular}
\caption{$F$-score, estimated file overlap, true overlap, and model-based overlap 95\% credible interval for the combinations of two models and two blocks for the Union Army data. $F$-score and estimated overlap are computed from the BRL and F-Algo point linkage estimates.}
\label{tab:union}
\end{table}

For each comparison vector set, we fit the Bayesian record linkage model described in Section \ref{sec:brl-primer} using the ``BRL'' package with default hyperparameters, collecting 20,000 posterior samples after a burn-in period of 5,000 iterations. Using the posterior samples, we compute and compare the BRL estimate from section \ref{sec:def_brl_estimator} and the {F-Algo} estimate from section \ref{sec:F-score-algo}.

Table \ref{tab:union} displays the results from the linkage with the Union Army data. In  the case of Block 1, the BRL and F-Algo estimates perform similarly, with the same $F$-score for model A and only a marginal difference for model B. For the estimated overlap, the F-Algo estimate is closer to the truth. In the case of Block 2, when there is more ambiguity, the F-Algo estimates have substantially higher $F$-scores than the BRL estimates. Furthermore, the estimated overlaps from the F-Algo estimate are inside the 95\% credible intervals, whereas the estimated overlaps for the BRL estimates are not. As seen previously, the F-Algo estimator tends to be less conservative than the BRL estimator in the presence of higher uncertainty, and thus can estimate the number of overlap links more accurately.

\section{Discussion}\label{sec:discussion}
We propose a post-processing algorithm for point estimation of the linkage structure in bipartite record linkage tasks. Given either a posterior distribution or pairwise match probabilities, the algorithm obtains a point estimate through approximately maximizing the expected $F$-score. The proposed optimization algorithm extends the approach of \citet{jansche-2007-maximum} to bipartite record linkage. By exploiting the sparsity of the linkage matrix, we implement several computational efficiency improvements, ensuring that the algorithm achieves a satisfactory complexity bound. Results from the simulation study and illustrative applications highlight the potential for improved performance when linking fields are measured with error compared to other estimators. 

One area for future research involves incorporating an explicit constraint for overlap size accuracy into the algorithm. Currently, optimizing over the expected F-score implicitly balances precision and recall. However, it would be possible to integrate this balance directly into the score function, penalizing deviations of the precision-recall ratio from unity. Implementing such a constraint may improve the algorithm's performance, particularly in scenarios with small overlap size, where our $F$-score optimization approach is currently less effective. Additionally,  while we focused on comparing $F$-score optimization with the BRL estimator, it also would be valuable to assess how the plug-in approach of \eqref{eq:bayesMS} compares to other methods, such as the frequentist approaches of \citet{fellegi:sunter} or \citet{Jaro1989}. Lastly, improvements can be made to enhance computational efficiency. A promising approach is the use of Bayesian or Lipschitz optimization algorithms to speed up the outer optimization step in \eqref{eq:bayes_est_optim}. By identifying and applying an appropriate Lipschitz bound, we can efficiently exclude values of $\hat{\mathbf{C}}_{\text{Bayes}}(k)$ from the search space that do not lead to global maximums. This strategy is especially valuable in problems with large $n_{\mathcal{B}}$, where the large discrete search space for the optimal $\hat{\mathbf{C}}_{\text{Bayes}}(k)$ may make it difficult to run the algorithm presented here efficiently.

\bibliographystyle{chicago}
\bibliography{biblio}

\appendix

\clearpage
\newpage

\end{document}